# Inverse Design of Conjugated Polymers from Computed Electronic Structure Properties: Model Chemistries of Polythiophenes

*Ilana Y. Kanal, Steven G. Owens, Andrey B. Sharapov, Geoffrey R. Hutchison\**

Department of Chemistry, University of Pittsburgh, 219 Parkman Avenue, Pittsburgh, PA 15260

geoffh@pitt.edu

**ABSTRACT.** Using a diverse set of 100 oligothiophenes, extrapolated oligomer and polymer HOMO, LUMO and HOMO-LUMO gaps are shown to be accurately estimated from the computed values of the trimer HOMO, trimer LUMO and trimer HOMO-LUMO gaps. These simple approximations can be improved further through easy to calculate properties of monomers or small oligomers. Polymer reorganization energies, related to the hole transport, are also shown to be predicted accurately from small oligomers. Correlations between the HOMO slope with the HOMO polymer energy and HOMO-LUMO gap slope with the polymer HOMO-LUMO gap energy suggest that the degree of delocalization reflected in the slope is higher in monomers and polymers with electron-rich, less negative HOMO eigenvalues. These statistical models can serve as efficient screens for the computational prediction of polythiophene properties.

**Introduction**

Polythiophene-derived materials continue to gather interest for both fundamental scientific study and technological applications, including field-effect transistors[1], light-emitting diodes[2], lithium-ion batteries[3], flexible electronic devices[4], and organic photovoltaics[5]. Arguably most of the important parameters for these applications derive from fundamental electronic structure parameters such as band gap, oxidation, and reduction potentials[6,7]. Consequently, optimizing these parameters is a key problem for the scientific community to solve.

The wide array of applications for polythiophene-based materials continues to drive both fundamental and application-based research. Many applications including field-effect transistors (FETs), light-emitting diodes (LEDs), batteries, flexible electronics, and organic photovoltaics (OPVs) have drawn considerable research interest. OPV applications of polythiophenes have become a major focus due to their easy processability and the effect relatively cheap photovoltaics would have on solar energy technology. The fundamental electronic structure of the molecule including properties like band gap and oxidation and reduction potentials play a large part in the properties of these materials. To better understand and improve the technological applications, these key underlying factors that impact the properties of the materials must be investigated.

There are tens of thousands of possible thiophene monomers with varying functional group substitution (electron donating/withdrawing and steric bulk), aromaticity/quinoidal character, and heteroatom substitution. These differences in monomer structure ultimately affect the electronic and structural properties of their polymers. Through a computational study of 100 thiophene monomers, we demonstrate how monomer properties determine the electronic and structural properties of the polymer. We seek to understand which key factors determine these important properties by investigating trends in properties such as HOMO, LUMO, band gap, dihedral angle, ligand width, van der Waals size of ligand, ligand type, average bond length, and bond length between carbon 3 and 4 which estimates aromaticity of the thiophene rings (**Scheme 1**).

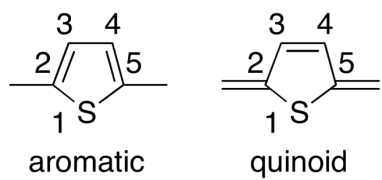

**Scheme 1.** Bond length between carbons 3 and 4 used to determine aromatic / quinoid character of a compound

By studying monomers and oligomers we can observe the relationship between monomer, oligomer, and ultimately polymer properties. The goal would be to trivially predict the properties of the extended polymer from the known or computed properties of the monomers – greatly reducing the computational and experimental time required. In the last several years, while large automatic searches and machine learning have become a major research focus,[8-19] [20] it is also important to carefully construct manual studies to test structure-property relationships.[21,22] In this study, we follow the later approach by examining a diverse set of polythiophenes, as shown in **Figure 1**, which shows a ~3 eV range for both HOMO and LUMO monomer and trimer values, including a range of both electron rich donors and electron poor acceptors. Our goal is to accurately predict properties of the homopolymers from monomers or small oligomers.

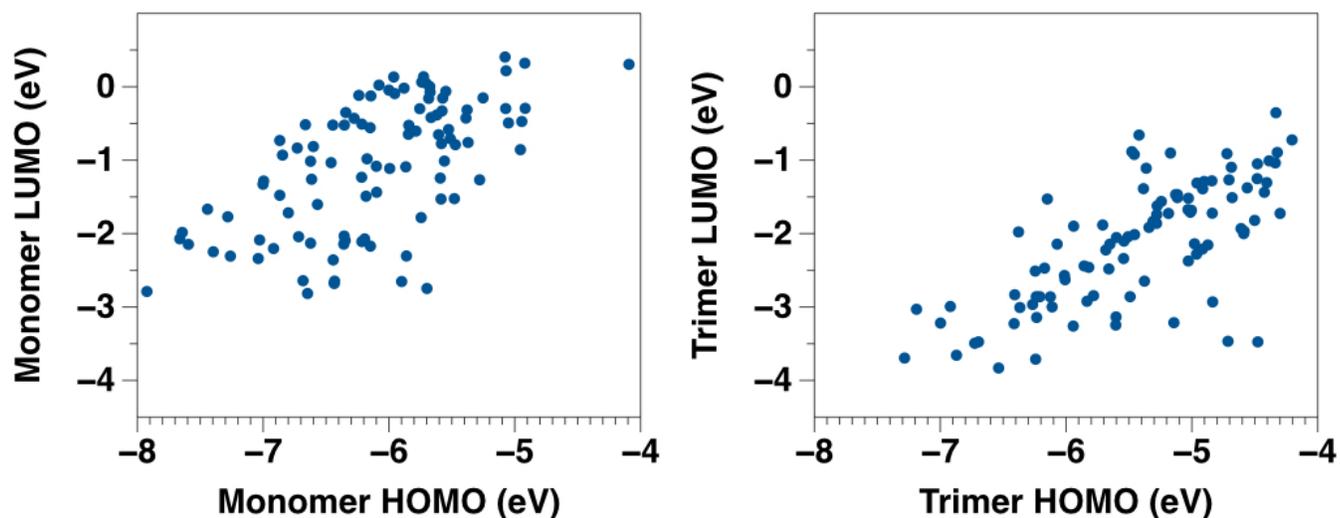

**Figure 1.** Monomer and trimer diversity demonstrated by a ~3 eV range in both the computed B3LYP HOMO and LUMO energy values. Note that both donor (electron rich, less negative HOMO/LUMO) and acceptor (electron poor, more negative HOMO/LUMO) monomers exist along a spectrum of properties.

II. **Computational Methods**

**Monomer Data Set**. The 100 monomers in this study were selected by composing a list of potential thiophene substitutions, including simple hand-picked functional groups, nonaromatic and aromatic fused rings. None of the chosen functional groups were sterically bulky (e.g. t-butyl, mesityl), to ensure relatively planar, conjugated oligothiophenes as a consistent set. For this study, monomers were limited to species containing C, H, N, O, S, Se, F, Cl, and Br. **Figure 2** shows the structures of the monomers studied and **Tables S1-S2** of the supporting information list the names of each of the monomers and literature references. SMILES are given in **Table S3**.

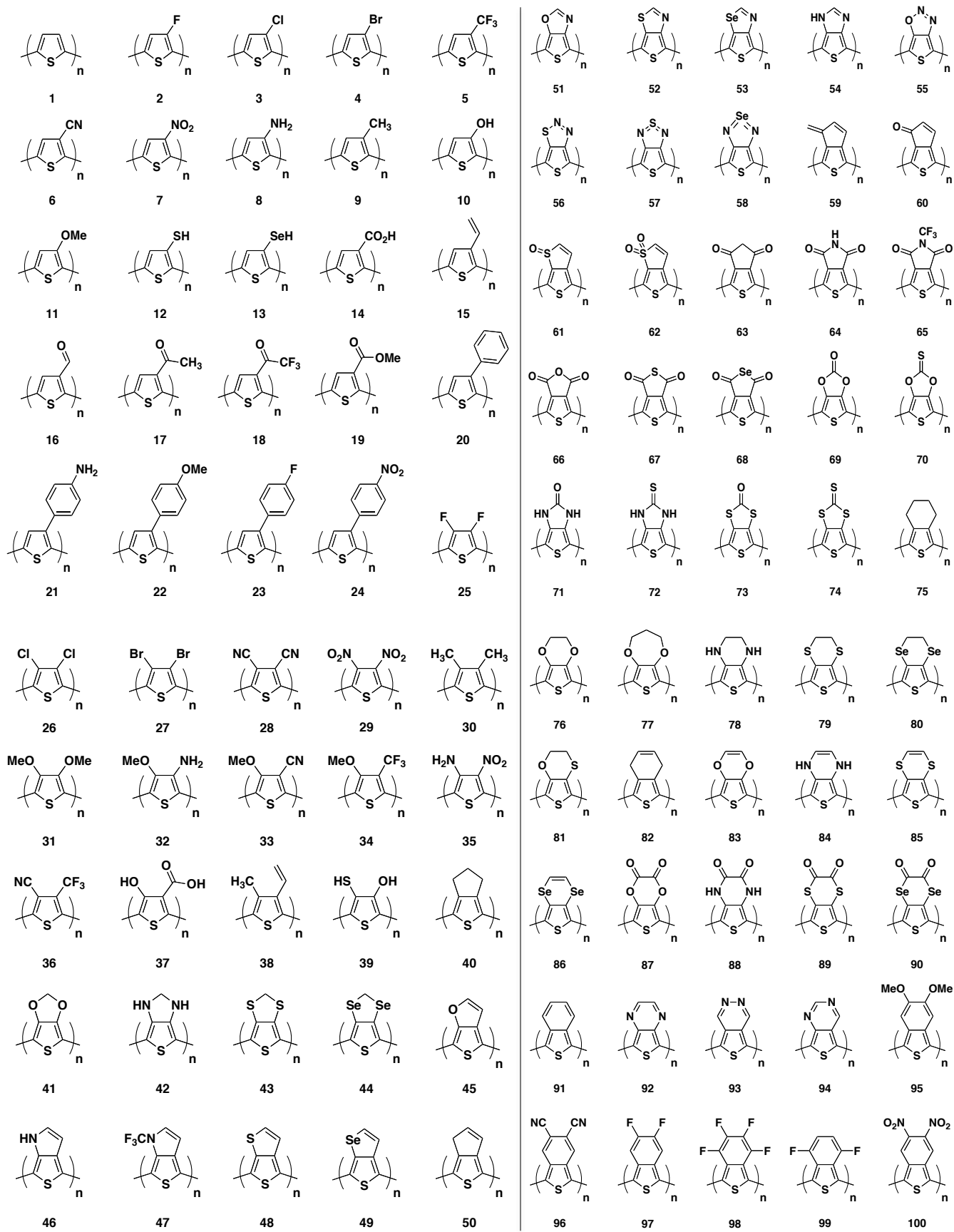

**Figure 2.** Chemical structure of the diverse oligothiophenes studied, including a range of mono- and di-substituted species, including a wide range of fused ring structures.

**Generation of Optimized 3D Structures.** For each oligomer, the 3D structure of a low-energy conformer was generated starting from the SMILES[23], generating the 3D structure using Open Babel[24] and minimized using the MMFF94 force field.[25-29] Next a weighted-rotor search (MMFF94, 100 iterations, 20 geometry optimization steps) was carried out with Open Babel to find a low-energy conformer. This was then further optimized using 500 steps of conjugate gradients and MMFF94. Finally, Gaussian09[30] was used to optimize the structure using DFT with the B3LYP[31,32] functional and the 6-31G* basis set. Although HOMO and LUMO eigenvalues from density functional methods cannot be formally taken as either the ionization potential or electron affinity, respectively, previous studies have shown that B3LYP-derived eigenvalues compare favorably with experimental electron affinities[33-36], ionization potentials,[33] and band gaps[37]. The idealized infinite polymer HOMO, LUMO, and HOMO-LUMO gap eigenvalues were determined based on linear regressions from the corresponding oligomer values versus the reciprocal oligomer length (i.e., $1/N$) [6,38,39].

**Statistical Methods.** As discussed above, this set of data contains 100 diverse thiophene monomers with different substitutions. The experimental goal of discovering a set of "descriptor properties" to be used to predict polymer properties by computing the monomer, or perhaps small oligomers such as dimer and trimer electronic structure, thus reducing computational time compared to the full polymer[40,41] was accomplished by performing applicable statistical tests and generate plots using R[42]. First, a set of distinct, unique, carefully chosen parameters were selected, including the values of elements 3 and 4, monomer ligand 3-4 width, element 3 van der Waals interaction, element 4 van der Waals interaction, ligand 3- 4 van der Waals interaction, charge of monomer 3, charge of monomer 4, monomer absolute charge, monomer average charge, HOMO, LUMO and HOMO-LUMO gap values of the monomer, dimer, trimer, tetramer and pentamer, dimer bond length alternation, tetramer bond length alternation, pentamer bond length alternation, monomer and pentamer length of the bond between elements 3 and 4, monomer and pentamer S charge, dihedral angles of the dimer, trimer, tetramer and pentamers. Care was taken to ensure that several parameters were not measuring similar properties that would skew the model results. A stepwise regression function, as discussed below, was used to predict the best set of predictors for the model by adding and subtracting different descriptors until the most predictive set was chosen without overfitting. Once this set of descriptors was chosen, it was defined as a particular model (e.g., examples described below). Still, many models that appear predictive, suffer from significant overfitting by incorporating too many descriptors.

To test for overfitting, each model was tested for reliability using both bootstrap and cross-validation tests. [43,44] Both of these methodologies provide information for how predictive a model will be when applied to new, unknown, data sets. Cross-validation is most reliable when studying large data sets since the data is divided into two parts, one for model development and one for model testing. In small data sets, each of these groups is not sufficiently large to allow for predictive model development. Hence, in our calculations, the bootstrap models consistently give better results since the data set only contains 100 species.

**III. Results**

Electronic properties were calculated for monomers and oligomers from two to five repeat units for all 100 thiophenes of interest in the study, and results were examined. Linear regressions and $R^2$ values for the infinite polymer HOMO, LUMO, and HOMO-LUMO gap values are shown in **Table S4**. The expected results from these calculations were negative HOMO slopes (i.e., increasing oligomer lengths yield less negative HOMO eigenvalues due to delocalization) and positive LUMO (i.e., increasing oligomer lengths yield more negative LUMO eigenvalues due to stabilized, delocalized electron affinities) and HOMO-LUMO gap slopes, as shown in **Figure 3**.

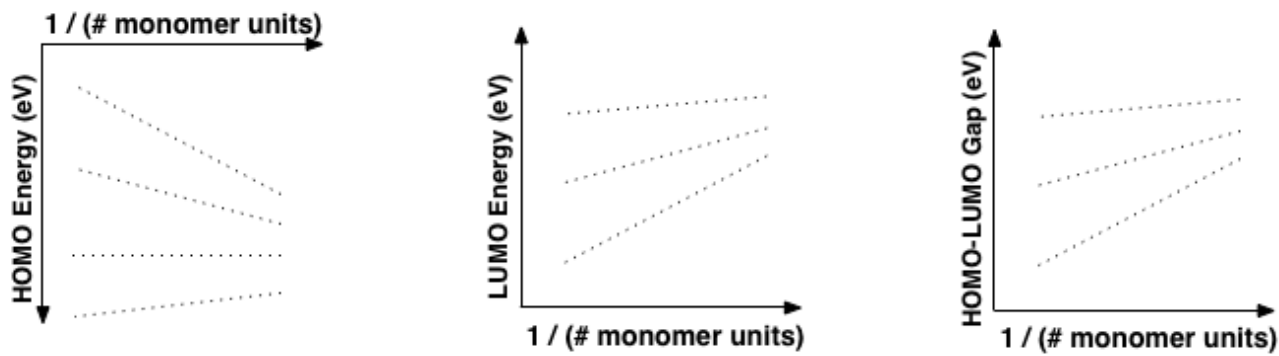

**Figure 3.** HOMO and LUMO expected energies as a function of the inverse of the number of monomer units, which demonstrates that the HOMO slopes are expected to be negative (i.e. increasing number of monomer units to the left results in higher HOMO energies) and LUMO slopes are expected to be positive.

A flat slope suggests that the electrons are not delocalized in a species (i.e., the electronic properties do not change as a function of oligomer length) while a high slope shows significant changes in the properties as the chain length changes. Most of the compounds follow these expected trends, with compounds **80**, **86**, **89**, **90** and **100** as exceptions to the trends. In addition, the $R^2$ values were expected to show high correlation which also was demonstrated, with the exception of compounds **26, 28, 29, 34, 44, 67, 80, 84, 86, 89, 90** and **100**, which show low $R^2$ statistics; while the properties are nearly linear, they are horizontal. Thus, calculations for these molecules suggest orbitals which are not delocalized, despite appearance as aromatic structures (and typical aromatic bond lengths, e.g., **Scheme 1**). The plots, therefore, appear to rise to a certain point and then flatten out as they switch from being aromatic to no longer being aromatic.

A natural question is what properties (e.g., geometric or chemical) alter the electronic properties such as HOMO, LUMO, and gap. Considering highly conjugated oligomers and polymers are typically planar, the average dihedral angle should have a significant effect on these properties. In **Figure 4**, we have graphed the effect of the steric crowding caused by the thiophene substituent, as measured by the distance between substituents (**Scheme 2**). Although as the chain size increases from dimer to pentamer, the $R^2$ increases from 0.32 to 0.44, there is no improvement when the chain size increases from the tetramer to the pentamer. Based on the modest $R^2$ values, torsional twisting caused by steric crowding does occur, but other effects also dictate dihedral angles in oligothiophenes. Moreover, we can see in **Figure 3**, that dimers do not yet reflect the dihedral angles of longer oligomers.

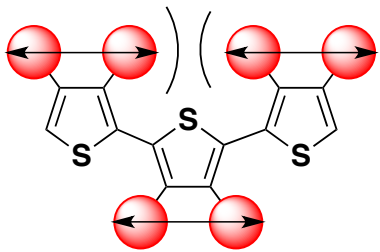

**Scheme 2.** Effect of steric crowding caused by the thiophene substituent "ligand width", as measured by the distance between substituents. (Black arrows)

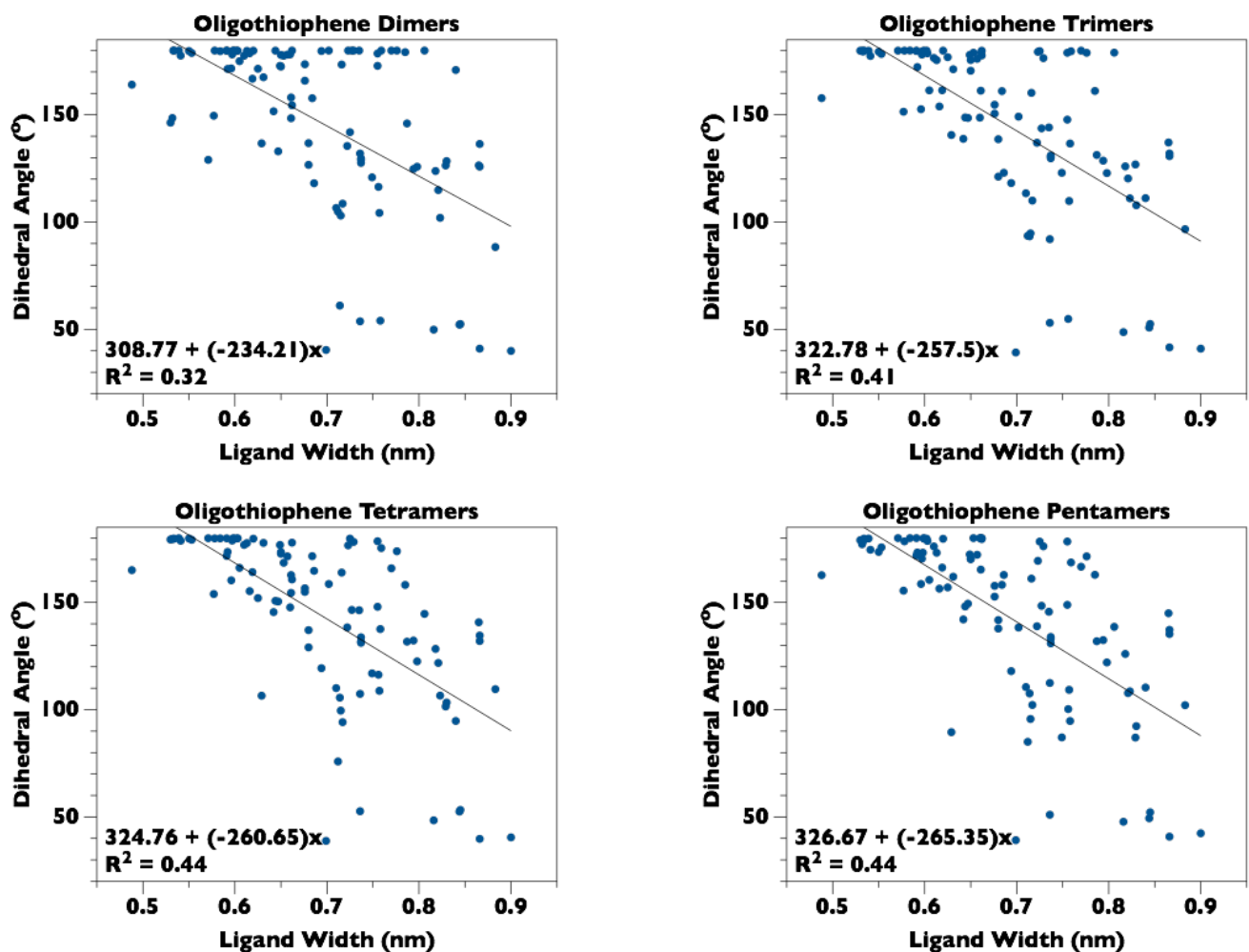

**Figure 4.** Ligand width correlation with the dihedral angle as shown with dimers, trimers, tetramers and pentamers.

**Computationally Efficient Models for Predicting Polymer Properties**

The main goal of this study is to determine if there is a simple way to predict the HOMO and LUMO values of a polymer from properties of the monomer, dimer, and trimer to create more efficient photovoltaics or other optoelectronic materials, by utilizing targets with a small optical gap and stable LUMO.

**HOMO**. The HOMO eigenvalues of monomers are only weakly correlated with the extrapolated HOMO energy of the infinite polymer, as summarized in **Table 1** (i.e., $R^2$ only 0.58 and mean unsigned errors of ~0.5 eV). Increasing to dimers significantly improves the correlation and decreases mean unsigned errors (MUE), but as shown in **Table 1, Figure 5** and elaborated in **Figure S1-2**, the trimer HOMO values reliably predict the polymer with $R^2$ of 0.94 and MUE < 0.2 eV. In order to ensure the significance of the model terms, an analysis of variance (ANOVA) test was run and the p-values show statistical significance in our models (i.e., $p < 2.2 \times 10^{-16}$ showing that the results are highly significant). In order to achieve lower mean absolute errors, models were prepared using other descriptors and stepwise regression to select optimal terms. The HOMO model combines the value of the monomer HOMO energy (MonHOMO), the trimer HOMO energy (3HOMO) and the value of the dihedral angle of the trimer (3Dihedral) as shown in the following equation:

$$\textbf{HOMO} = 0.82 - 0.28(\textbf{MonHOMO}) + 1.44(\textbf{3HOMO}) + 0.002(\textbf{3Dihedral}) \qquad (1)$$

This model achieves slightly higher predictability as demonstrated by the HOMO $R^2$ value increasing from 0.94 to 0.97, while lowering MUE to 0.11 eV, which is less than the known error from DFT (~0.2 eV).

**Table 1:** Summary of statistical correlation and predictive power for HOMO, LUMO and HOMO-LUMO gap models

| Model | Adjusted $R^2$ | Mean Unsigned Errors (eV) |
|---|---|---|
| **HOMO ~ Monomer HOMO** | 0.58 | 0.54 |
| **HOMO ~ Dimer HOMO** | 0.87 | 0.29 |
| **HOMO ~ Trimer HOMO** | 0.94 | 0.19 |
| **HOMO ~ HOMO Model** | 0.97 | 0.11 |
| LUMO ~ Monomer LUMO | 0.74 | 0.36 |
| LUMO ~ Dimer LUMO | 0.89 | 0.24 |
| LUMO ~ Trimer LUMO | 0.93 | 0.18 |
| LUMO ~ LUMO Model | 0.96 | 0.12 |
| Gap ~ Monomer Gap | 0.11 | 0.66 |
| Gap ~ Dimer Gap | 0.53 | 0.49 |
| Gap ~ Trimer Gap | 0.77 | 0.33 |
| Gap ~ Gap Model | 0.87 | 0.23 |

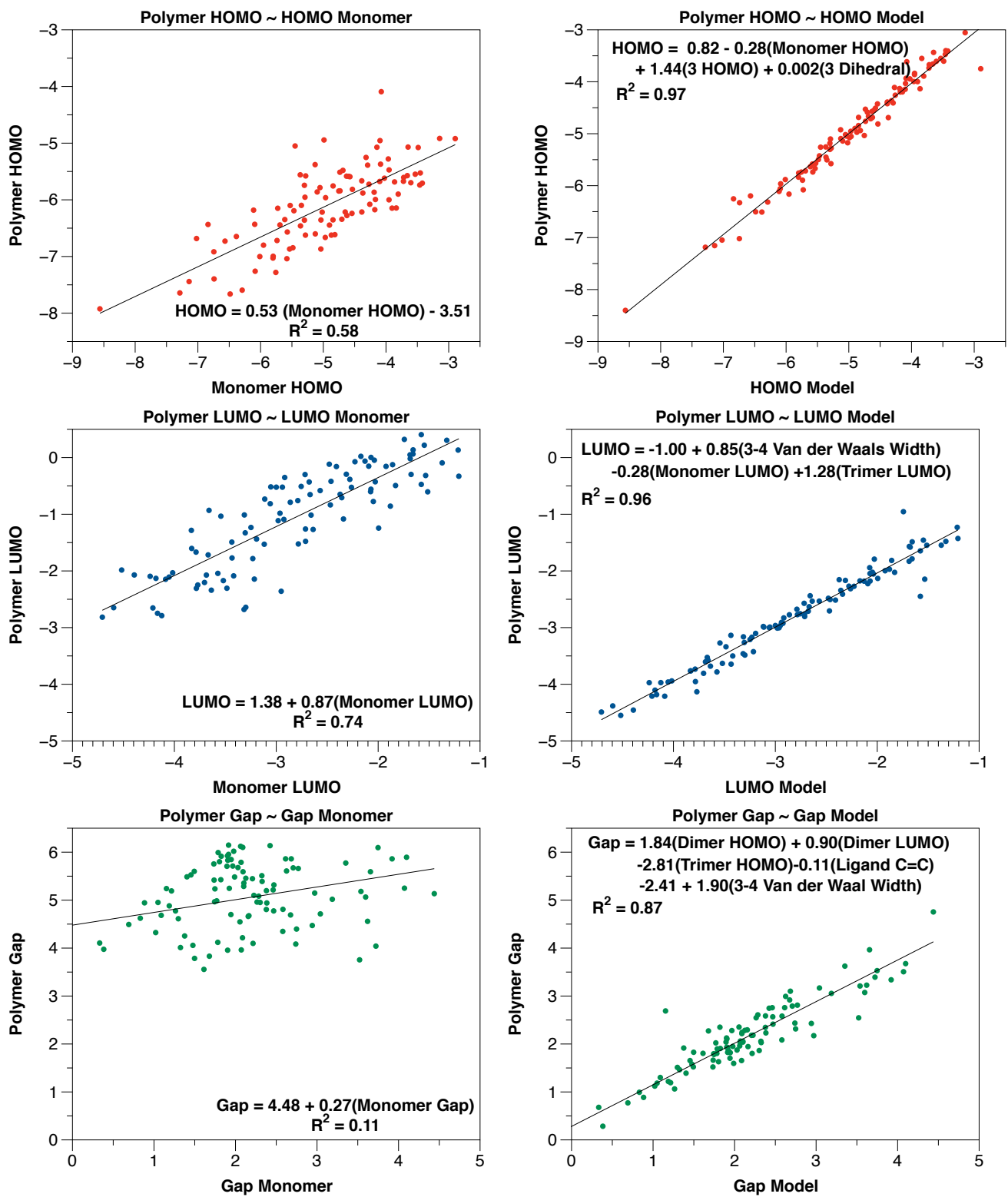

**Figure 5:** Linear regression models for predicting polymer HOMO, LUMO, and HOMO-LUMO gap from monomers (left column) and multivariate fits.

**LUMO**. Much like the HOMO eigenvalues, the monomer LUMO energies are modestly correlated with the polymer LUMO ($R^2$ 0.74) As shown in **Table 1, Figure 5** and elaborated on in **Figure S5**, the trimer LUMO values reliably predict the polymer with $R^2$ 0.93 and MUE under 0.2 eV. From the ANOVA test, the p-values for the trimer LUMO as a predictor of the polymer LUMO are $2.2 \times 10^{-16}$, showing that the results are statistically significant. An improved LUMO model combines the values of the monomer and trimer LUMO energies with the van der Waals width of the molecule in the following equation:

$$\mathbf{LUMO} = -1.00 - 0.28(\mathbf{MonLUMO}) + 1.28(\mathbf{3LUMO}) + 0.85(\mathbf{vdW\ Width}) \qquad (2)$$

As with the HOMO case, this increases the probability of correctly calculating the polymer LUMO energy since the $R^2$ increases from 0.93 to 0.96, and MUE ~0.12 eV. Including the "steric width" of the substituents likely indicates some effect of the dihedral angle on the LUMO energies.

**HOMO-LUMO Gap**. Despite the accuracy of models for the HOMO and LUMO, the HOMO-LUMO band gap correlations were surprisingly less accurate. Unlike the HOMO and LUMO monomer energies, which demonstrated some correlation with the polymer HOMO and LUMO energies, the gap of the monomer shows very low correlation with the band gap of the polymer as indicated by the $R^2$ value of 0.11. The low $R^2$ value can be attributed to the slopes of the HOMO and LUMO energies not correlating to one another as shown through their respective coefficients. The band gap of the trimer shows improvement over the monomer by increasing the $R^2$ to 0.77 and decreasing MUE to ~0.3 eV. After deriving multivariate models for the HOMO and LUMO polymer energies, the expectation was that the trimer gap would be selected in the model for the gap, as the HOMO and LUMO trimer energies are part of the HOMO and LUMO polymer models, respectively. The most predictive model for the HOMO-LUMO gap of the polymer, instead combines the dimer HOMO and LUMO energies, the trimer HOMO energy, the 3-4 van der Waals width and if the ligand attached to the molecule has a C=C (**Scheme 3**):

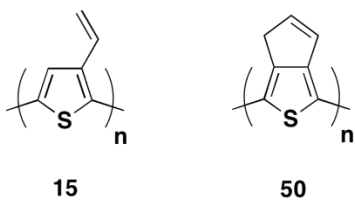

**Scheme 3.** Examples of substituents with carbon-carbon double bonds

$$\mathbf{Gap} = -2.41 + 1.84(\mathbf{2HOMO}) + 0.90(\mathbf{2LUMO}) - 2.81(\mathbf{3HOMO}) \\ + 1.90(\mathbf{vdW\ Width}) - 0.11(\mathbf{Ligand\ C=C}) \qquad (3)$$

This equation shows improvement by increasing the $R^2$ to 0.87 and decreasing MUE to ~0.2 eV (**Table 1**). In conclusion, the most predictive multivariate models involve the HOMO and LUMO trimer energies, and the correlation increases when other descriptors are added to the HOMO and LUMO trimer energies to slightly increase the predictability of the model.

**Reorganization Energies.** Beyond orbital energies and HOMO-LUMO gap values, an important factor in organic electronics are the molecular Marcus reorganization energy. [6,7] Much like the orbital eigenvalues, one might seek a rapid predictor of polymer reorganization energies on the basis of monomer or small oligomer properties. A similar trend to the one described for prediction of polymer HOMO, LUMO, and HOMO-LUMO gap, as described above was found for the reorganization energies (**Table S5**). Low correlation is shown from the monomer λ compared with higher correlation from dimer or trimer reorganization energies. This encouraged a continued search to discover an accurate multivariate model to predict the reorganization energy. The dihedral angle and bond length alternation were found to have little correlation with the reorganization energy, as verified by a model with an $R^2$ of 0.22. The reorganization energy also shows little correlation with the polymer HOMO or LUMO, even though the HOMO and LUMO slopes are indicators of the polymer energies and delocalization. The final model:

$$\lambda_{\text{Pentamer}} = -0.018 + 0.259(\lambda_{\text{Trimer}}) + 0.681(\lambda_{\text{Tetramer}}) + 0.023(\text{Ligand C=C}) \quad (4)$$

combines the lambda of the trimer and tetramer with whether the ligand has a carbon-carbon double bond for slightly better correlation (0.78 to 0.89) and MUE < 0.05 eV. (**Table 2, Figure S3**)

**Table 2.** Summary of models for internal reorganization energies for hole transport

|  | Adjusted $R^2$ | Mean Unsigned Error (eV) |
|---|---|---|
| λ Pentamer ~ λ Monomer | 0.09 | 0.16 |
| λ Pentamer ~ λ Dimer | 0.43 | 0.11 |
| λ Pentamer ~ λ Trimer | 0.78 | 0.06 |
| λ Pentamer ~ λ Model | 0.89 | 0.04 |

During hole transport, the reorganization energy reflects an activation barrier to charge transfer. Consequently, large reorganization energy can be considered as a "filter" for organic electronic materials. That is, oligomers or polymers with high computed reorganization energies are not likely to have high charge mobility. Compounds **5, 16, 19, 22, 24, 26, 27, 29, 30, 31, 33, 34, 35, 36, 38, 47, 55, 68, 75, 79, 82, 88,** and **89** (**Figure S4**) were found to have large computed hole reorganization energies and are not ideal targets. The following trends that emerge suggest that large, bulky groups decrease mobility. Monomer **5** compared with similar monomers (**1-4** and **6-14**) suggests that the $CF_3$ group with its three fluorine atoms decreases the mobility compared with the other monomers which have less bulky substituents. Monomers **22** and **24** compared to similar monomers (**20, 21, 23**) indicates that a thiophene with a fused benzene substituent loses mobility when the benzene is substituted with OMe and $NO_2$. Monomers **26** and **27** when compared with monomer **25** show that chlorine and bromine, show lower mobility, potentially due to larger steric bulk compared to fluorine, producing a more highly twisted oligomer backbone.

On the other hand, compounds **4, 6, 40, 45, 46, 48-54, 56-58, 60, 80, 92** and **94** (**Figure S5**) were found to small reorganization energies (< 0.07 eV) and are likely to have high relative mobilities and are good candidates for efficient charge transport. Although few trends are obvious, monomer **40** compared with similar monomers **41-44** suggests that nonaromatic five-membered rings made solely from carbon, have much higher activation energy than heterocycles with oxygen, nitrogen, sulfur or selenium.

Finally, we find multiple targets with highly negative LUMO eigenvalues, likely to be good electron acceptors (i.e., high electron affinity). Compounds **28, 29, 55, 56, 57, 58, 65, 66, 93, 94, 96**, and **100** (**Figure S6**) all show highly negative LUMO energies. A clear trend is the combination of aromatic substituents with two nitrogens (eg. oxadiazoles, thiadiazoles, diazines, etc.) which strongly stabilizes the LUMO orbital.

IV.  Discussion

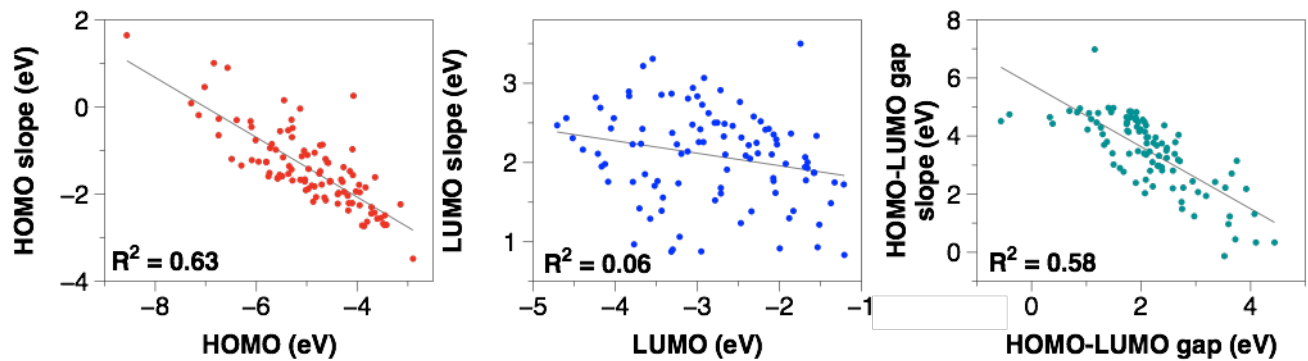

**Figure 6.** Correlations between HOMO, LUMO and HOMO-LUMO gap slopes with the predicted polymer energies.

An important question from the results, is the surprising success of the "one point" correlations, for example, accurately predicting the extrapolated polymer HOMO solely from the trimer or tetramer value. Since an infinite number of lines can be fit through one point, (e.g., **Figure 2**), it seems unlikely that only one point is needed. The results indicate, however, a high correlation ($R^2$ of 0.63) between the slope of the linear regression in HOMO eigenvalues and the extrapolated polymer HOMO (i.e., the y-intercept). A similarly high correlation ($R^2$ of 0.58) is found for the slope of the HOMO-LUMO gap and the extrapolated gap (**Figure 6**). Such correlations suggest that the degree of delocalization reflected in the slope is higher in monomers and polymers with electron-rich, less negative HOMO eigenvalues. Indeed, we find that the slope is *positive* for monomer **29**, dinitrothiophene; for this repeat unit, longer oligomers have *more negative* HOMO eigenvalues than the monomer, suggesting oligo-dinitrothiophene would be harder to oxidize than the monomer. These correlations, for the slope of HOMO eigenvalues and HOMO-LUMO gaps, suggest that electron-rich donor monomers are predicted to have high delocalization and, in general, low HOMO-LUMO gap for the homopolymers.[45]

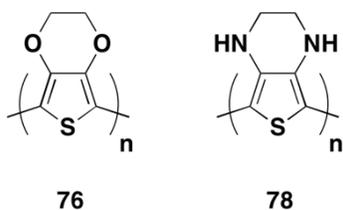

**Scheme 4.** PEDOT (76) versus PDAT (78)

The results indicate several interesting applications. For example, PEDOT, which is widely used, has very similar properties to PDAT (78), including similar band gap and HOMO energies (**Scheme 4**). The trimer HOMO energy for PEDOT (-4.34 eV) and the trimer HOMO energy for PEDAT (-4.33 eV) are very similar, predicting that these will have similarly easy oxidation. Additionally, the extrapolated polymer band gap of these two are similar, suggesting PDAT warrants further investigation.

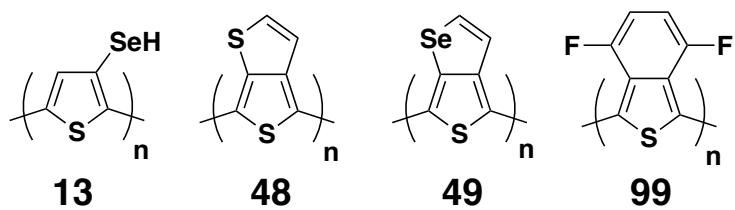

**Scheme 5.** Polymers with predicted OPV device efficiencies > 8% by Scharber criteria

While most materials for efficient organic photovoltaics use complex "donor-acceptor" co-polymers, we note that several polythiophenes homopolymers studied here (**Scheme 5**) are predicted to have efficiencies above 8%, based on the Scharber criteria.[46] Notably, all four have high (less negative) HOMO energies, suggesting a donor-donor strategy may provide an alternative to the current donor-acceptor design.

Most efforts in the field to improve OPV efficiency have focused on the p-type polymer, but to improve power efficiencies, all properties of the solar cell must be considered and improved. Substituted fullerenes such as PCBM has been n-type phases with wide successes in OPV devices.[47] To improve efficiency, changing to a different n-type material with similar LUMO energy might prove worthwhile, by giving stronger optical absorption. Monomers **28**, **29**, **55-58**, **65-66**, **93-94**, **96** and **100** (**Figure S6**) have LUMO values between -4-5 eV making them potential replacements for PCBM or other fullerene acceptors.

To further test the HOMO, LUMO and gap models, a similar thiophene, outside the initial set of 100 species, with an electron drawing cyano group substitution (**Scheme 6**), was considered.

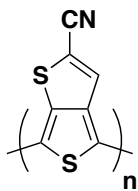

**Scheme 6. Test Monomer**

For this compound, as expected from our previous discussion, the slope of the LUMO versus 1/N is a positive value (+2.57 eV/[# repeat units]), the slope of the HOMO versus 1/N is a negative value (-0.85 eV/[# repeat units]), and the HOMO-LUMO gap slope is positive (+3.41 eV/[# repeat units]). In addition, the values were tested using the equations of the trimer models shown above to verify that the model is able to accurately predict HOMO and LUMO values from novel monomers. The actual HOMO value of -5.33 eV and the value of the trimer HOMO equation of -5.24 eV and the actual LUMO value of -3.89 eV and the value from the trimer LUMO equation of -3.69 eV show that the calculations performed with the model are accurate within ~0.2 eV.

V.    **Conclusions**

We have shown that across a set of 100 diverse oligothiophene species, the polymer HOMO, LUMO and HOMO-LUMO gaps, computed from DFT, can be accurately estimated from the values of the trimer HOMO, trimer LUMO and trimer HOMO-LUMO gap calculated values. We also show that these approximations can be improved through the models presented above, including simple and easy to calculate properties. In all three cases, the resulting mean unsigned errors are at or below ~0.2 eV, well within the error of the computational method. Consequently, rather than performing multiple oligomer calculations to extrapolate the polymer electronic structure, most properties can be estimated readily from modest-sized oligomers.

Polymer reorganization energies, related to the hole transport, are also shown to be predicted accurately from small oligomers. The pentamer reorganization energies, as representatives for the polymer, are shown to be highly correlated with the trimer reorganization energy, but this correlation and accuracy increase with other descriptors.

We speculate that the accuracy of the "one point" model, based on the properties of the trimer, occur because this length approximates the dihedral angles of the polymer based on steric crowding between ligands and other effects. Moreover, while an infinite number of possible lines could fit through one point, we find the slope of the regression lines are correlated. That is, species with high, less negative HOMO energies, show greater shifts as a function of oligomer length. Moreover, strong acceptor monomers, such as 3,4-dinitrothiophene **29**, show an unusual oligomer slope, in which dimers and oligomers are predicted to be harder to oxidize (more negative HOMO) than the monomer.

Our overall trends related to individual molecules did not yield surprising results, but show that large, bulky groups reduce mobility as the polymer chain increases in length and a more highly twisted oligomer backbone emerges. The high correlation between the slope of the linear regression in HOMO eigenvalues and the extrapolated polymer HOMO and analogous comparison with the HOMO-LUMO gap suggest that the degree of delocalization reflected in the slope is higher in monomers and polymers with electron-rich, less negative HOMO eigenvalues. Additionally, we have demonstrated that the amine analogue to PEDOT, with its similar chemical electronic structure requires further attention from the community. Finally, we find that four homopolymers in our sample group yield predicted OPV device efficiencies > 8% by Scharber criteria, have high (less negative) HOMO energies, suggesting a donor-donor strategy may provide an alternative to the current donor-acceptor design.

Models, such as those presented here, can be used in larger projects exploring chemical space as reliable ways to predict polymer properties from smaller molecules such as trimers and tetramers. We believe that these statistical models can serve as rapid "first screens" for a wide range of optoelectronic electronic structure properties.

**Acknowledgement**. We thank the University of Pittsburgh, including the Center for Simulation and Modeling for computational resources, Pitt Center for Energy for support, and NSF (CBET-1404591). GRH thanks Dr. Noel O'Boyle and Prof. Tara Meyer for discussions.

**Supporting Information**. IUPAC names, references, and SMILES for thiophene molecules studied, scatterplot of parameters studied, additional graphs demonstrating HOMO, LUMO and HOMO-LUMO gap trends, and schematics of molecules discussed in the discussion are included as supporting information. This material is available free of charge via the Internet at http://pubs.acs.org and at http://hutchison.chem.pitt.edu

**Table of Contents Graphic:**

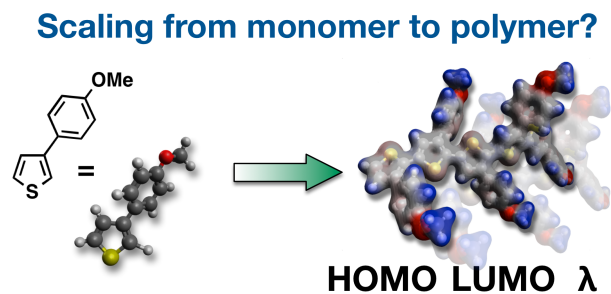